\documentclass{ws-mpla}

\begin{document}
\def\draftnote{Preprint: IP/BBSR/2003-41}

\markboth{P. Arumugam, T.K. Jha \&  S.K. Patra} {Role of BCS-Type
Pairing in Light Deformed Nuclei...}

%%%%%%%%%%%%%%%%%%%%% Publisher's Area please ignore %%%%%%%%%%%%%%%
%
\catchline{}{}{}{}{}
%
%%%%%%%%%%%%%%%%%%%%%%%%%%%%%%%%%%%%%%%%%%%%%%%%%%%%%%%%%%%%%%%%%%%%

\title{ROLE OF BCS-TYPE PAIRING IN LIGHT DEFORMED NUCLEI: A RELATIVISTIC
MEAN FIELD APPROACH}

\author{P. Arumugam$^1$, T.K. Jha$^2$ and  S.K. Patra$^1$}

\address{$^1$Institute of Physics, Bhubaneswar - 751 005, India \\
$^{2}$ P.G. Department of Physics, Sambalpur University,
Jyotivihar, Burla - 768 019, India}

\maketitle

\pub{Received 14 November 2003} {}%{Revised (Day Month Year)}

\begin{abstract}
We calculate the binding energy and deformation parameter for
light nuclei {\it with} and {\it without} pairing using a deformed
relativistic mean field model. The role of BCS-type pairing effect
is analyzed for Ne and F isotopes. The calculated odd-even
staggering and the deformation parameters argue strongly against
the role of pairing in the light nuclei.

\end{abstract}

\ccode{PACS Nos.: 21.10.Re, 21.30.Fe, 21.60.Jz, 27.70.+q}

\vspace*{12pt}

The odd-even staggering (OES) of nuclear masses has been
recognized since the early days of nuclear physics. It manifests
itself in the fact that the binding energy of a system with an odd
particle number is lower than the arithmetic mean of the energies
of the two neighboring even-particle-number systems. H\"akkinen
{\it et al.} \cite{hakk1}, using the density-functional theory,
argued that light alkali-metal clusters and light $N=Z$ nuclei
have a similar pattern of OES, irrespective of differences in the
interactions between the fermions. Hence, they concluded that the
OES in small nuclei appears to be a mere deformation effect rather
than a consequence of pairing.

On the other hand, Satula {\it et al.} \cite{satu1}, claimed that
the OES in light atomic nuclei is strongly affected by both
nucleonic pairing and the deformed mean field. According to them,
the OES without taking pairing into account is simply due to
Jahn-Teller effect \cite{jahn1} and it is substantially smaller
than the experimental observation. It is worthwhile to note that,
following Ref. \cite{bohr69}, the pairing energy is about $-0.8$
MeV for a nuclear system of mass number $A=160$. The rather small
value of the correlation energy reflects the fact that only a few
single particle levels lie within the interval of strong pair
correlations. Thus the condition for the treatment in terms of a
static pair field is only marginally satisfied. This makes clear
that the concept of pairing may have some meaning for heavy mass
nuclei. But the basis for pairing correlations in light nuclei is
very questionable.

The Bohr-Mottelson-Pines idea of applying BCS type pairing theory to nuclei
and the Mottelson-Valatin model of pairing disappearance
\cite{bohr58} are attractive ideas from the theoretical
point of view, one should still approach this problem with caution, since
in a nucleus one is dealing with a small number of particles. The number
of particles and the number of states available near the Fermi
surface are too limited in a nucleus, whereas in a solid (where the
BCS theory originated) one has an enormously large number
of particles and number of states.

Taking into consideration the above contradictory predictions, in the
present letter we calculate the OES with and without pairing for
Ne-isotopes. We
obtain the binding energies and other quantities of interest by
performing axially deformed relativistic mean field (RMF)
calculation \cite{patra1,ring1}.
We show that the inclusion of BCS-type pairing does not improve the
OES for light nuclei. In addition, the calculated result of deformation
parameter argues strongly against pairing.

We start with the relativistic Lagrangian of interacting nucleons
and mesons (protons, neutrons, scalar and vector mesons along with
electromagnetic field interacting with charged protons). The
Euler-Lagrange field equations for mesons, nucleons and
electromagnetic potentials are obtained \cite{patra1,ring1}. The
nucleons satisfy the Dirac equations in the potential field of the
Bosons. The mesons and the photons are assumed to be classical
fields whose sources are the densities and currents of nucleons.
These equations are solved by expanding the upper and lower
components of the Dirac spinors and the Boson field wave functions
in axially symmetric deformed harmonic oscillator basis with some
initial deformation. A large number of oscillator shells
($N_{F}=N_{B}=12$) are used as the basis for expansion of the
Fermion and Boson fields. The set of coupled equations are solved
numerically by self-consistent iteration method with the NL3
parameter set \cite{lala1}. It is to be noted that the RMF
parameter sets are determined by fitting the nuclear matter
properties, neutron-proton asymmetry energy, root mean square
radii and binding energy of some spherical nuclei. There are no
further adjustments of parameters to be made. Hence any
discrepancy with pairing model for the nuclei is a genuine result.
The quadrupole deformation parameter $\beta_2$ is evaluated from
the resulting quadrupole moment.

\begin{figure}[th]
\centerline{\psfig{file=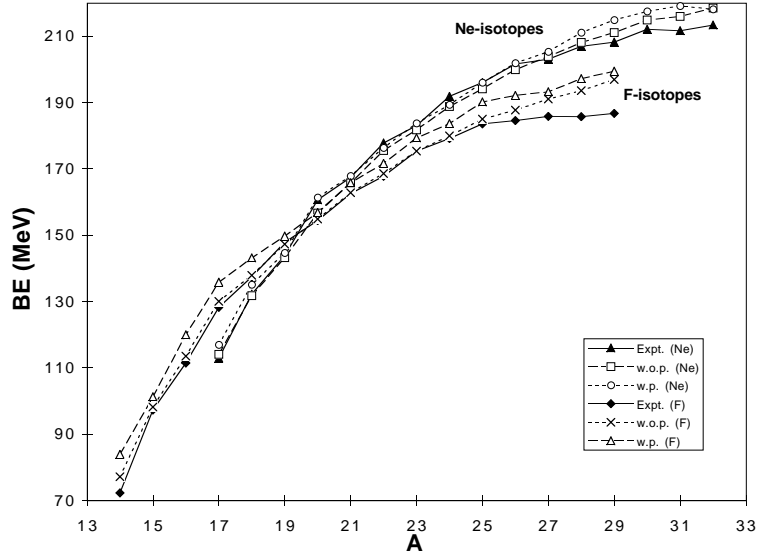,width=0.78\textwidth,clip=true}}
\caption{The calculated binding for the RMF theory,
using NL3 parameter set versus mass number A for Ne and
F-isotopes. The experimental (or extrapolated) data are shown for
comparison.}
\end{figure}

We do RMF calculations with and without pairing. The pairing
calculations are done in the constant gap approximation. The
constant gaps for protons and neutrons are evaluated from the
equations \cite{bohr1}:
\begin{equation}
\triangle_{n}={1\over 4}\{BE(N-2,Z)-3BE(N-1,Z)+3BE(N,Z)-BE(N+1,Z)\}
\end{equation}
\noindent and
\begin{equation}
\triangle_{p}={1\over 4}\{BE(N,Z-2)-3BE(N,Z-1)+3BE(N,Z)-BE(N,Z+1)\}
\end{equation}
where $BE(N,Z)$ is the experimental (or extrapolated from data)
binding energy \cite{audi1} with neutron number $N$ and proton
number $Z$. When the experimental (or extrapolated) data is not
know (e.g., for nuclei away from stability valley), the gap is
evaluated from the general expression of Bohr and Mottelson
\cite{bohr1}:
\begin{equation}
\triangle \approx {12\over{A^{1/2}}}  {\rm MeV}
\end{equation}
The calculations for the odd-even and odd-odd nuclei in an axially
symmetric deformed basis is a tough task in the RMF model. To take
care of the lone odd nucleon, one has to violate the time-reversal
symmetry. However, in the present calculation we have not
carefully taken into consideration such effects. At the level of
bulk properties for nuclei, the effect of the time reversal
symmetry is not very significant, while its effect is significant
in the determination of other (time-odd) properties like the
magnetic moment, etc. \cite{hofm1}. Therefore, it is expected that
the results for binding energy and quadrupole moment may not
change much when one performs the exact calculation by including
the effects of the time-reversal asymmetry for the odd nucleons in
an odd-even or odd-odd system. In the present RMF study, apart
from the $\sigma$ meson field, only the time components $V_{0}$,
$\rho_{0}$, $A_{0}$ of the $\omega$, $\rho$ and photon fields are
retained, since these are important for the ground state bulk
properties. A measurable time-asymmetric field would find many
nucleons in time-reversal-asymmetric configurations and hence
constitutes an excited state. Hence the space components of these
fields (which are odd under time-reversal and parity), as also the
time-reversal breaking effect of odd nucleon, are not considered
for the bulk properties (binding energy and matter distribution).
Thus the odd nucleon is considered for its contribution to binding
energy and matter distribution in the approximation of a
time-reversal symmetric mean-field (average of the $\pm m$ lone
nucleons are taken for the mean-field potential). Observables like
binding energy, charge or matter density and quadrupole moment
correspond to even operators as far as time-reversal symmetry is
concerned and hence are fairly well explained in the approximation
scheme adopted by us.

\begin{figure}[t]
\centerline{\psfig{file=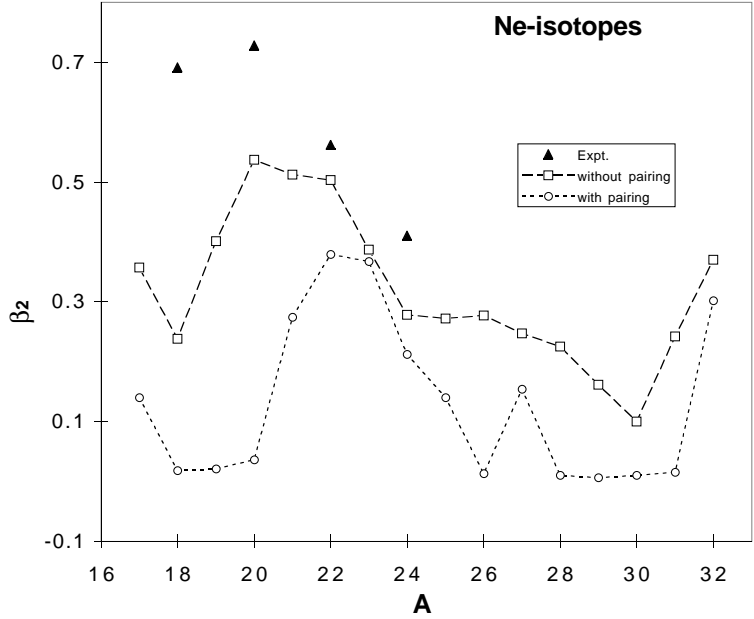,width=0.78\textwidth,clip=true}}
\caption{Plots of quadrupole deformation parameter
versus mass number $A$, {\it with} and {\it without} pairing,
using NL3 parameter set. The solid triangles are the experimental
data.}
\end{figure}

Let us now come to the results of our RMF calculations. First of
all we calculate the binding energy ($BE$) and quadrupole
deformation parameter ($\beta_2$) for Ne and F isotopes, setting
$\triangle_{p}=\triangle_{n}=0$.  From the calculated binding
energies we evaluate the one-neutron ($S_{n}$) and one-proton
separation $S_{p}$ energies using the relations \cite{bohr1}:
\begin{equation}
S_{n}(N,Z)=BE(N,Z)-BE(N-1,Z)
\end{equation}
and
\begin{equation}
S_{p}(N,Z)=BE(N,Z)-BE(N,Z-1)
\end{equation}
Again we repeat the calculations taking $\triangle_n$ and
$\triangle_p$ into account as defined in equations (1-3) and
evaluated the values of $S_n$ and $S_p$ using equations (4) and
(5), respectively.

The binding energy with and without pairing for Ne and F isotopes
are compared with experimental data \cite{audi1} in Fig. 1. From
the figure, it is clear that the inclusion of pairing does not
improve the binding to any significant extent. On the other hand
the results without pairing are favored for the majority of the
isotopes. In case of F-isotopes, the inclusion of pairing gives
severe over binding. In Fig. 2, we have compared the quadrupole
deformation parameter ($\beta_2$) with the experimental
measurements \cite{raman1}. Here the $\beta_2$ values are
calculated {\it with} and {\it without} taking pairing into
consideration. The $\beta_2$ values without pairing (square) gives
better fit with experimental measurements for known cases, whereas
the RMF calculation predicts very small quadrupole deformation
parameter in the presence of pairing. In an earlier paper
\cite{patra3}, we found that the binding energy ($BE$) and root
mean square radii are not very sensitive to the pairing
correlation, but the shapes of the nuclei, particularly near the
stability line, depend very much on the pairing gap. A comparison
with experimental data shows that a zero pairing strength is
preferred in the light and light-medium nuclei region.

It is well-known that even-even nuclei have consistently more
binding than neighboring even-odd (or odd-even) nuclei. The
binding energies of odd-odd nuclei are systematically less than
these. This is due to binding energy gain arising from interaction
$V_{pp}$ or $V_{nn}$ for nucleons occupying time-reversal
symmetric orbits ($\pm m$ degenerate orbits in axially symmetric
nuclei) in even particle configurations.  Usually BCS or
Hartree-Fock-Bogoliubov (HFB) methods are used to study pairing in
nuclei \cite{bohr69,bohr58,lane64,baranger62,pal82}. A closer
study shows that odd-even staggering is present in the binding
energies of deformed nuclei in a mean field description.

\begin{figure}[t]
\centerline{\psfig{file=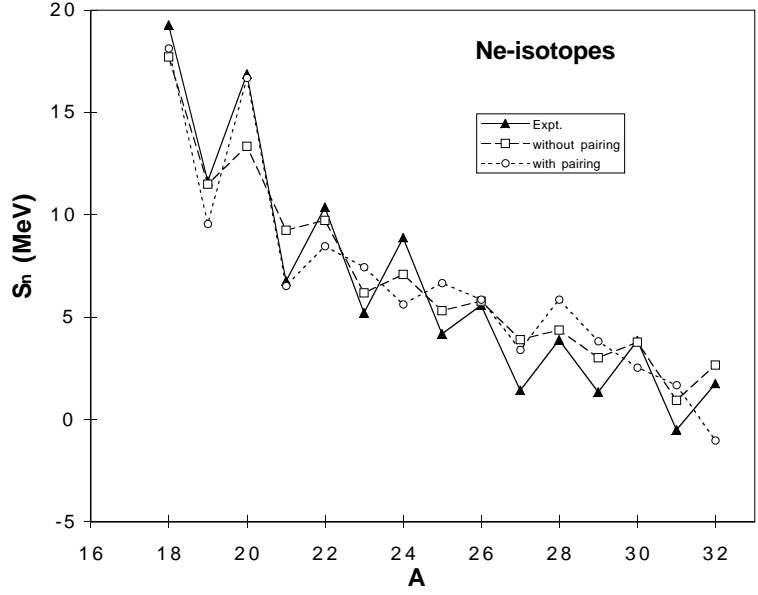,width=0.78\textwidth,clip=true}}
\caption{The calculated one-neutron separation
energy ($S_n$) for the RMF calculations compared with the
experimental data for Ne-isotopes. }
\end{figure}

The one-neutron ($S_n$) and one-proton ($S_p$) separation energies
for Ne-isotopes are shown in Fig. 3(a) and Fig. 3(b),
respectively. From Fig. 3(a) it is clearly evident that the $S_n$
values fit well with pairing  for $A=19$ and 20, whereas for all
other isotopes the one-neutron separation energy is closer to the
experiment in the absence of pairing (square in the figure).
Similarly the $S_p$ values argue against pairing. The calculated
results are closer to the experimental data for $S_p(10,10)$ and
$S_p(11,10)$ with pairing, whereas it very much deviates (remains
same with the {\it without} pairing case for $S_p(18,10)$,
$S_p(19,10)$ and $S_p(20,10)$) from the experimental data for rest
of the isotopes (see Fig. 3(b)). The quadrupole deformation for
$^{27,28,29}$F are almost identical in both {\it with} and {\it
without} pairing cases. Thus the OES is also found to be almost
same. The value of OES, as also the magnitude of $\beta_2$,
assumes different values in cases {\it with} and {\it without}
pairing.

The OES is caused by interaction of nucleons occupying {\it
paired} orbitals in even nuclei and absence of such {\it paired}
partner for the lone nucleon in odd nucleus. Thus interactions
$V_{pp}$ and $V_{nn}$ and configuration mixing present in deformed
mean field is the major factor governing OES in deformed nuclei
\cite{praharaj98,praharaj94}. No BCS or HFB pairing (with the
attendant number fluctuations) is needed for deformed nuclei. It
is to be noted here that the quadrupole deformation parameter
becomes worse, when we include the BCS-type pairing correlation
(Fig. 2). Also the OES gets affected. On the other hand, in the
absence of pairing, the $\beta_2$ values as well as the OES are
reproduced quite well. This indicates that the odd-even staggering
in binding energy is mainly due to the effect of the mean field
(not due to pairing).

\begin{figure}[t]
\centerline{\psfig{file=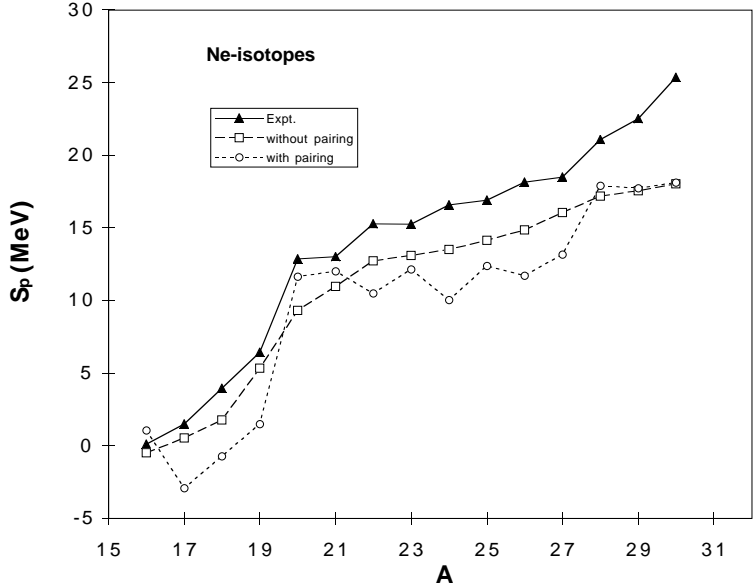,width=0.78\textwidth,clip=true}}
\caption{Same as Fig. 3. but one-proton separation
energies ($S_p$) are plotted here.}
\end{figure}

In summary, we calculated the binding energy and quadrupole
deformation parameter in the frame-work of RMF formalism for
Ne-isotopes. We analyzed the effects of BCS-type pairing
correlation on odd-even staggering and $\beta_2$. From the present
RMF results (Figs. 1-3), we draw the following conclusions:
\begin{romanlist}[(ii)]

 \item The inclusion of pairing does not change the binding energy
drastically for light nuclei region. From a further inspection of
the binding energy, it is observed that the calculated binding
energies are closer to the experimental data in the absence of
BCS-type pairing correlation.

  \item The trend of quadrupole deformation parameter
rules out the presence of BCS-type pairing in light nuclei region.

  \item The comparison of $S_n$ and $S_p$ with the experimental
values clearly shows that the odd-even staggering in light nuclei
is a mean field phenomenon.
\end{romanlist}

We thus conclude that BCS (or HFB) type pairing correlation is not the
cause for the odd-even staggering in light nuclei. It is due to the
Jahn-Teller mean field effects.

\section*{Acknowledgments}

One of the authors (TKJ) thank the Institute of Physics, Bhubaneswar,
for providing research facilities.

\end{document}